\documentclass[a4paper]{ESASPCS13Style}
\usepackage{epsfig}

\newcommand{\cmsq}{cm$^{-2}$}
\newcommand{\cmcc}{cm$^{-3}$}

\newcommand{\kms}{km~s$^{-1}$}

\newcommand{\ergps}{erg~s$^{-1}$}
\newcommand{\flx}{ph~cm$^{-2}$~s$^{-1}$}

\newcommand{\chan}{\textit{Chandra}}

\begin{document}

\title{High-Energy Processes in Young Stars:
\chan\     X-Ray Spectroscopy of 
HDE 283572, RY Tau, and L\lowercase{k}C\lowercase{a} 21}

\author{M.~Audard\inst{1} \and S.~L. Skinner\inst{2} \and K.~W. Smith\inst{3} \and
M. G\"udel\inst{4} \and R. Pallavicini\inst{5}}
  
  \institute{Columbia Astrophysics Laboratory, Columbia University, Mail code 5247, 550 West 120th
  Street, New York, NY 10027, USA \and
  Center for Astrophysics and Space Astronomy, University of Colorado, Boulder,
  CO 80309-0389, USA \and
  Max-Planck-Institut f\"ur Radioastronomie, Auf dem H\"ugel 69, 53121
  Bonn, Germany \and 
  Paul Scherrer Institut, Villigen \& W\"urenlingen, 5232 Villigen PSI,
  Switzerland \and 
  Osservatorio Astronomico di Palermo, Piazza del Parlamento 1, 90134
Palermo, Italy}

\maketitle 

\begin{abstract}
Weak-lined T Tauri stars (WTTS) represent the important stage of stellar evolution 
between the accretion phase and the zero-age main sequence. At this stage, the star 
decouples from its accretion disk, and spins up to a higher rotation 
rate than in the preceding classical T Tauri phase. Consequently, dynamo processes 
can be expected to become even stronger at this stage. High energy processes can have 
effects on the remaining circumstellar material, possibly including protoplanets 
and planetesimals, and these effects may account for certain observable properties 
of asteroids in the current solar system. \chan\   observed 
for 100 ks the WTTS HDE 283572 which probes the PMS stage of massive A-type 
stars. We present first results of the analysis of its high-resolution X-ray 
spectrum obtained with the High-Energy Transmission Grating Spectrometer. A wide 
range of Fe lines of high ionization states are observed, indicating a continuous 
emission measure distribution. No significant signal is detected longward of the 
O~\textsc{viii} Ly$\alpha$ line because of the high photoelectric absorption. We also report on 
the preliminary analysis of the zeroth order spectra of RY Tau and LkCa21. 
In particular, we show evidence of an emission line in RY Tau at 6.4 keV that we 
identify as fluorescent emission by neutral Fe caused by a strong X-ray flare 
which illuminated some structure in (or surrounding) the CTTS. A comparison of X-ray spectra of 
classical T Tau stars, other WTTS, and young main-sequence stars is made.

\end{abstract}

\section{Introduction}
Weak-lined T Tauri stars (WTTS) are of special interest for the study of magnetic
activity in the early stages of a star. WTTS are pre-main sequence (PMS) stars
between the accreting phase of a classical T Tau star (CTTS) and the disk-free
phase of a zero-age main sequence (ZAMS) star. While the X-ray properties of 
protostars and CTTS may be related to the phenomenon of accretion, 
X-rays in WTTS are thought to originate from magnetic 
activity due to a dynamo mechanism similar to main-sequence magnetically active stars. 
While accretion disks in CTTS keep the rotation rate low, the highest rotation rates 
occur in WTTS, due to the dispersion of the disk. Magnetic activity is thus very strong in these 
PMS stars ($10^{28.5} - 10^{31}$~\ergps). WTTS are, therefore, most 
important for our deeper understanding 
of extreme magnetic activity in young stars, of the evolution of the
circumstellar envelope and disk, and of the formation of planetary systems (e.g., \cite{feigelson99}).

The origin of X-ray emission in young PMS stars still remains unclear.
X-ray grating spectroscopy of accretion-fed CTTS have shown stimulating,
but conflicting results: \object{TW Hya}, a nearby CTTS seen pole-on, showed
a soft X-ray spectrum (3~MK) with high densities ($\log n_\mathrm{e} \sim
13$~\cmcc; \cite{kastner02}; \cite{stelzer04}), whereas another
CTTS, \object{SU Aur}, displayed a spectrum characteristic of a dominant very hot plasma
($T \sim 30-50$~MK; \cite{skinner98}; \cite{smith05}).
Whereas accretion has been claimed as the mechanism at the origin of X-rays in TW
Hya, enhanced solar-like coronal activity appears to explain the X-ray spectrum
of SU Aur. Although X-ray studies of CTTS are interesting, the lack of
significant accretion in WTTS makes the latter excellent candidates to study magnetic 
activity in PMS stars.

\section{Targets and Observations}
\label{sec:obs}

\chan\  observed the WTTS star \object{HDE 283572} with ACIS-S and the High-Energy Transmission 
Grating for 102~ks from October 20, 2003 (14h45m17s UT) to October 21, 2003 (19h37m43s UT).
The nearby ($8.5\arcmin$ separation) WTTS \object{LkCa 21} was detected in the zeroth order image
only, whereas the
CTTS \object{RY Tau} was bright enough to produce significant signal in the grating arms as well.
Table~\ref{tab:targets} gives some properties of the targets.
 
\begin{table}[!bht]
  \caption{Target properties.}
  \label{tab:targets}
  \begin{center}
    \leavevmode
    \footnotesize
    \begin{tabular}[h]{llll}
      \hline \\[-5pt]
      {} 			&HDE 283572    		&  HD 283571     	&  LkCa 21  \\[+5pt]
      \hline \\[-5pt]
      A.k.a.     		& V987 Tau		& RY Tau  		& V1071 Tau	\\
      SED class 		& III			& II      		& III 		\\
      Spectral type		& G5 IV$^a$		& F8 III$^b$   		& M3$^c$ 	\\
      Age (Myr) 		& $3^c$			& $1.6^d$		& $0.5^e$ 	\\
      Mass ($M_\odot$) 		& $1.8\pm 0.2^a$ 	& $1.4^f$		& $0.2^e$  	\\
      $v \sin i$ (km/s) 	& $78 \pm 1^a$ 		& $55 \pm 3^b$ 		& $60 \pm 11^g$	\\ 
      $P_\mathrm{rot}$ (d) 		& $1.55^{a,h}$ 		& $\dots$		& $\dots$	\\
      $A_V$ (mag) 		& $0.4-0.6^a$  		& $2.7^i$  		& $0.65^c$	\\
   $\log N_\mathrm{H}$~()	& $20.60^j$ 		& $21.70^j$ 		& $< 21.3^j$	\\
   $\log L_\mathrm{X}$ (erg/s)	& $31.14^j$  		& $30.88^j$ 		& $29.83^j$ 	 \\
      \hline \\[-5pt]
      \multicolumn{4}{l}{\parbox{0.95\hsize}{References:
      $^a$\cite*{strassmeier98}, %
      $^b$\cite*{mora01}, %
      $^c$\cite*{damiani95}, %
      $^d$\cite*{hartigan95}, %
      $^e$\cite*{martin94}, %
      $^f$\cite*{basri91}, %
      $^g$\cite*{hartmann87}, %
      $^h$ \cite*{walter87},%
      $^i$ \cite*{beckwith90},%
      $^j$ This work   }}
      \end{tabular}
  \end{center}
\end{table}

\begin{figure}[!ht]
\centering
\resizebox{\hsize}{!}{\includegraphics{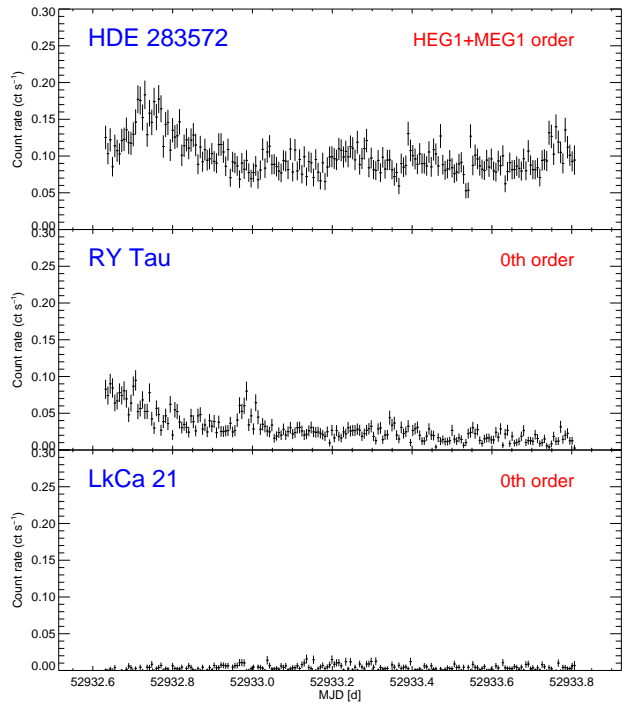}}
\caption{\chan\  X-ray light curves of HDE 283572, RY Tau, and LkCa 21 (top, middle, and bottom panels).
An identical vertical scale was used in each panel to emphasize the different X-ray fluxes.
The count rate in
the grating light curve of HDE 283572 is similar to that in the zeroth order.
\label{fig:lc}}
\end{figure}

Our targets lie in the Taurus-Auriga star-forming region at $d \sim 140$~pc.
The main target, HDE 283572, was discovered in X-rays and studied in detail
by \cite*{walter87} and later by \cite*{favata98a}. Its age and 
spectral type suggest that it is a probable predecessor 
of an A-type main-sequence star (\cite{favata98b}). 
It rotates fast ($v \sin i \sim 80$~\kms; \cite{johnskrull96}) and it  is probably 
observed equator-on, implying a short rotation period (\cite{walter87}). 
In X-rays, HDE 283572 displays a dominant hot plasma ($10-40$~MK) and
its light curves show evidence for rapid and intense flare-like 
variations (\cite{favata98a}; \cite{strom94}; \cite{stelzer00}). 
Both RY Tau and LkCa 21 lie about $8\arcmin$ north of HDE 283572 and were previously detected
in X-rays (\cite{damiani95}). Their X-ray fluxes are such that their count rates are about twenty
times lower than that of HDE 283572. RY Tau was caught during the decay of a flare, and produced
sufficient signal in the grating arms. However, the large off-axis distance significantly distorted the
redistribution matrix thus degrading the spectral resolution of the grating spectra. Figure~\ref{fig:lc} shows
the X-ray light curves in the HETGS first order (HDE 283572) and in zeroth order (RY Tau and LkCa 21).

\section{Data Reduction}
\label{sec:datared}

The HETGS data were reduced with CIAO 3.0.2 and CALDB 2.26. We followed threads to produce a calibrated
event type 2 file. In particular, we took care that events of RY Tau and LkCa 21 were properly
calibrated. For  HDE 283572, we assumed that no other contaminating source exists. While,
in principle, order sorting could have differentiated crossing grating events of RY Tau (MEG $+1$) from those of HDE
283572 (HEG $-1$), this was not possible for this observation since both grating arms
crossed at about the same energy. Thus, in the analysis below, we have discarded the $10.8-11.6$~\AA\  range
in HEG $-1$.

\begin{figure}[!ht]
\centering
\resizebox{\hsize}{!}{\includegraphics{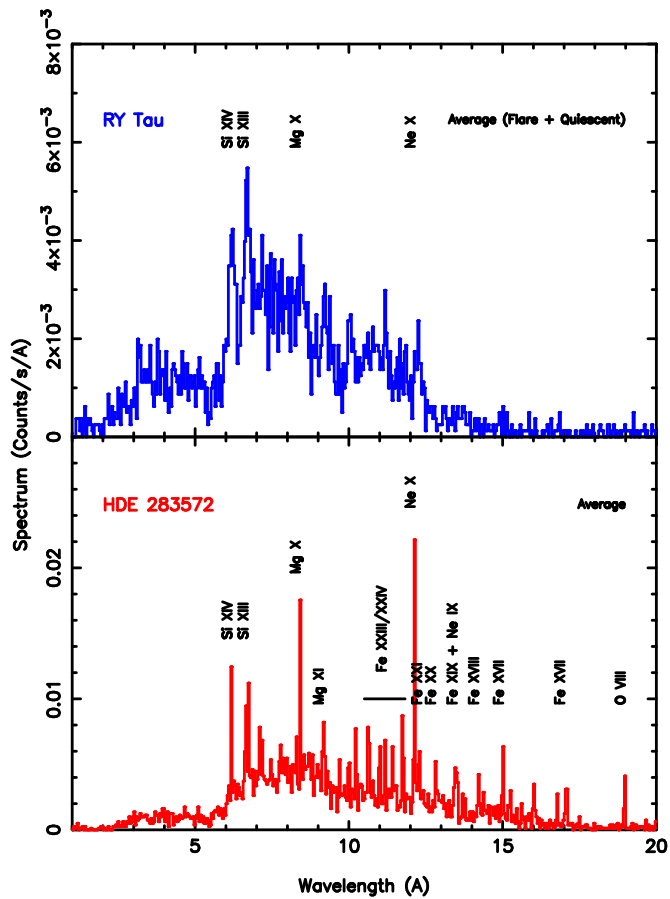}}
\caption{\chan\  MEG first order average spectra of RY Tau (top) and HDE 283572 (bottom) with major
emission lines labeled. Notice the 
strong continuum in both hot sources but the higher photoelectric absorption in
the CTTS. The resolving power
of the grating spectrum is also reduced at off-axis angles.\label{fig:grating}}
\end{figure}

\begin{figure}[!ht]
\centering
\resizebox{\hsize}{!}{\includegraphics[angle=-90]{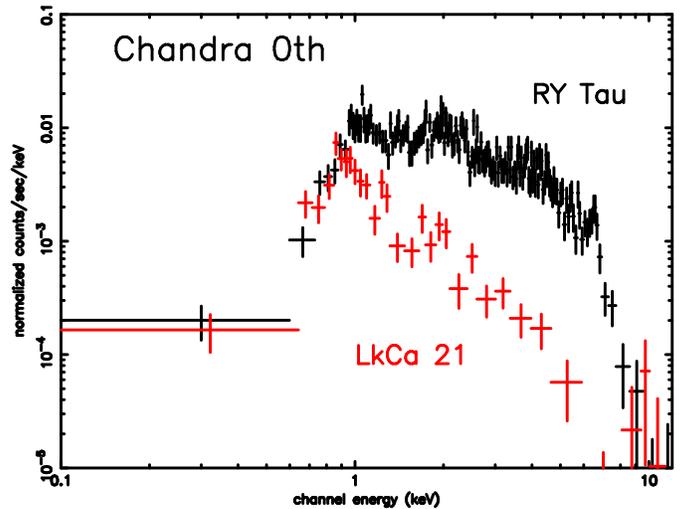}}
\caption{\chan\  zeroth order average spectra of RY Tau (top) and LkCa 21 (bottom). Notice the similar
photoelectric absorption at low energy but much hotter spectrum of RY Tau.
\label{fig:zeroth}}
\end{figure}

\section{Spectra}
\label{sec:spec}

Figures~\ref{fig:grating} and \ref{fig:zeroth} show the \chan\  MEG first order spectra of RY Tau and HDE
283572, and the zeroth order spectra of RY Tau and LkCa 21, respectively. A few features can immediately
be recognized by eye: i) the spectra of HDE 283572 and RY Tau both show a well-developed continuum at
short wavelength, indicating a dominant hot plasma, with RY Tau being slightly hotter. ii) However, photoelectric absorption is much
stronger in RY Tau than in HDE 283572. iii) A similar hydrogen column density can be found in the WTTS
LkCa 21 and in the CTTS RY Tau. iv) The plasma in LkCa 21 is much cooler than in
the  other two targets. 
We present below preliminary results of our data analysis. Specifically,
uncertainties are calculated as 20\% of the best-fit values only; furthermore, whereas the grating
spectra of HDE 283572 were analyzed, we fitted only the zeroth order spectra 
of RY Tau and LkCa21.
Solar photospheric abundances from \cite*{grevesse98} are used here.

\subsection{HDE 283572}

We fitted simultaneously the HEG and MEG first positive and negative order spectra after rebinning to
obtain a similar bin size of 20~m\AA. The negligible background was not subtracted allowing us to use the
robust C statistics. We found that two absorbed isothermal collisional ionization equilibrium models
could fit the data adequately. The best-fit parameters (with 20\% quoted uncertainties) are
$T_1 = 8.5 \pm 1.7$~MK, $T_2 = 23 \pm 4.6$~MK, $\mathrm{EM_1} = (3.3 \pm 0.7) \times 10^{53}$~\cmcc\  and 
$\mathrm{EM_1 / EM_2} = 0.6$ ($\log L_\mathrm{X} = 31.14$~\ergps; $0.1-10$~keV). We here fixed $N_\mathrm{H}$ to $4 \times 10^{20}$~\cmsq, i.e., the ROSAT 
PSPC value \cite*{favata98a}. This value is consistent with that obtained from a spectral
fit to the zeroth order spectrum of HDE 283572. Coronal abundances were obtained as well: $\mathrm{O} =
0.48 \pm 0.1$, $\mathrm{Ne} = 0.80 \pm 0.16$, $\mathrm{Mg} = 0.41 \pm 0.08$, $\mathrm{Si} = 0.35 \pm
0.07$, $\mathrm{S} = 0.18 \pm 0.04$, $\mathrm{Ca} = 1.12 \pm 0.22$, $\mathrm{Ar} = 0.94 \pm 0.19$,
and $\mathrm{Fe} = 0.41 \pm 0.08$. No additional component was required (C statistics of 3711 for 
3230 degrees of freedom). However, this does not exclude the presence of a cooler
component in HDE 283572. The photoelectric absorption in HDE 283572 prevented the detection of
cooler plasma that could, e.g., be detected in the O~\textsc{vii} triplet.

\subsection{RY Tau}

\begin{figure}[!t]
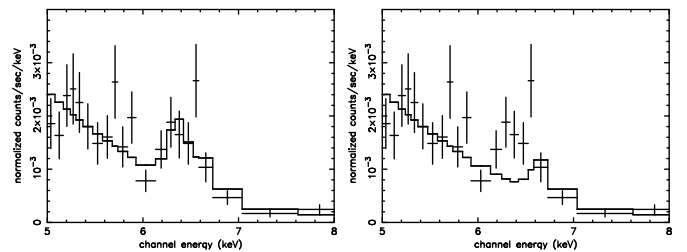

\centering
\resizebox{\hsize}{!}{\includegraphics[angle=-90]{maudard2_4a.eps}\hspace*{5mm}
\includegraphics[angle=-90]{maudard2_4b.eps}}
\caption{Zoom-in of the zeroth order spectrum of RY Tau around the Fe K line complex. An additional
line component is required to fit a flux excess  observed at 6.4~keV. A model with
(left) and without (right) the emission line is shown.\label{fig:rytaufluo}}
\vskip -2mm
\end{figure}

In this proceedings paper, we concentrate our analysis on the zeroth order 
spectrum of RY Tau. 
A 2-$T$ model fit yields $T_1 = 12 \pm 2.4$~MK, $T_2 = 45 \pm 9$~MK, 
$\mathrm{EM_1} = (1.9 \pm 0.4) \times 10^{53}$~\cmcc\  and 
$\mathrm{EM_1 / EM_2} = 0.5$ ($\log L_\mathrm{X} = 30.88$~\ergps; $0.1-10$~keV). We obtained a global metallicity of $Z = 0.3 \pm 0.06$ and $N_\mathrm{H} =
(5 \pm 1) \times 10^{21}$~\cmsq. Most interestingly, an excess flux around 6.4~keV was observed in RY
Tau. An additional emission feature with flux of $(4.3 \pm 1.9) \times 10^{-6}$~\flx\ 
(90\% confidence range) was required. Figure~\ref{fig:rytaufluo} shows the high-energy region of the RY
Tau spectrum with (left) or without (right) the additional component. This emission feature is reminiscent of
fluorescent emission by Fe ions due to inner shell ionization of a 1$s$ electron, and has been observed
in the Sun (e.g., \cite{neupert67}) and in young stellar objects (YSOs;
\cite{imanishi01}). The energy resolution of the
ACIS detector does not allow us to determine the exact ionization state of the emitting ion;
however, the line centroid ($6.37 \pm 0.05$~keV) suggests that Fe~\textsc{ii} (originally Fe~\textsc{i}
before K-shell ionization), e.g., in the accretion disk, could well have fluoresced.

\subsection{LkCa 21}

LkCa 21 was too faint for grating spectroscopy. A 2-$T$ fit to the zeroth order spectrum yields $T_1 = 
10 \pm 2$~MK, $T_2 = 25 \pm 5$~MK, $\mathrm{EM_1} = (0.35 \pm 0.07) \times 10^{53}$~\cmcc\  and 
$\mathrm{EM_1 / EM_2} = 1.2$, together with a global metallicity of $Z = 0.4 \pm 0.08$ and $N_\mathrm{H}
\leq 2 \times 10^{21}$~\cmsq\  ($\log L_\mathrm{X} = 29.83$~\ergps; $0.1-10$~keV).  
Thus this X-ray faint WTTS shows much cooler plasma than the brighter HDE 283572.

\section{Discussion}
\label{sec:discussion}

\begin{figure}[!t]
\centering
\resizebox{\hsize}{!}{\includegraphics[angle=0]{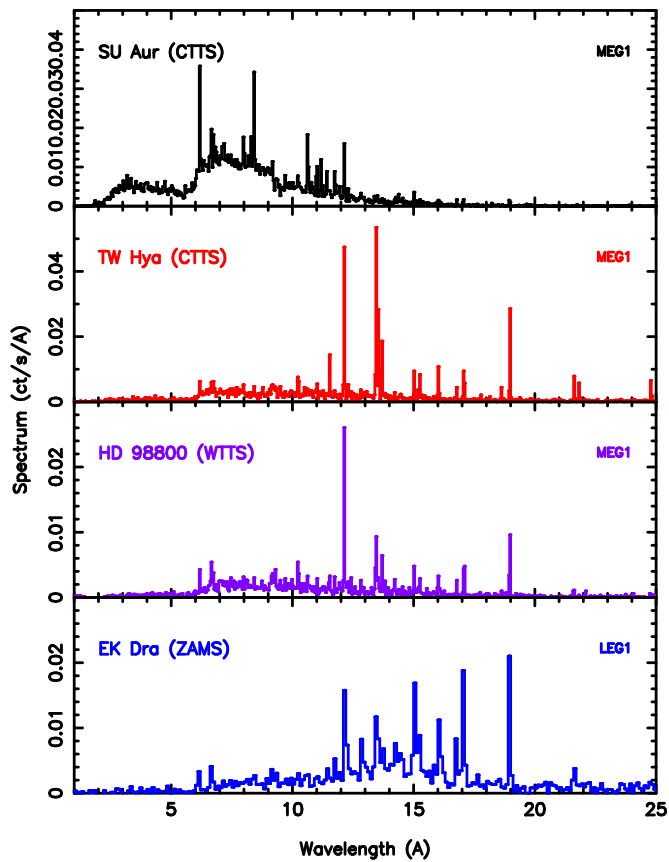}}
\caption{A sample of \chan\  X-ray spectra of CTTS (SU Aur and TW Hya), WTTS (HD 98800), and ZAMS star (EK Dra).
\label{fig:compare}}
\vskip -5mm
\end{figure}

Figure~\ref{fig:compare} shows the \chan\  X-ray spectra of CTTS SU Aur (age $\sim 5 $~Myr; 
\cite{smith05}), CTTS TW Hya (\cite{kastner02}), 
WTTS HD 98800 (\cite{kastner04}), and ZAMS star EK Dra (\cite{telleschi05}). 
During the accretion phase, the CTTS in our sample display very hot plasmas ($\sim 30$~MK), 
with the exception of TW Hya (age $\sim 5-10$~Myr) which
shows cool plasma ($3$~MK). Again different spectra can be observed at the WTTS stage: whereas the
corona of HD 98800 is cool ($<10$~MK), HDE 283572's spectrum is similar to the hot CTTS. The ZAMS star EK
Dra (age $\sim 50-70$~Myr) also shows a dominant cool plasma. In summary, no obvious trend can be seen
from Class II to III in this small sample. 

The detection of a fluorescence line in RY Tau may be explained by the flare
observed in the X-ray
light curve. It suggests that the X-ray flare source illuminated a cold structure, e.g., the accretion
disk in RY Tau, and produced fluorescent emission after K-shell ionization by the hard X-ray photons.
Further analysis will allow us to obtain some geometric constraints on the height of the X-ray source
above the fluorescing material.

In conclusion, some CTTS display strong levels of magnetic activity; however, it remains unclear where TW Hya,
because of its cool plasma and high electron densities,  fits into this picture. Similarly, coronal
emission measure distributions in WTTS are not unique. It remains unclear what properties dictate the X-ray emission
in YSOs. Could it be age, rotation, stellar association, external conditions, etc.
Hopefully, additional high-resolution X-ray spectra of PMS stars will shed light on this fundamental problem.

\begin{acknowledgements}
We acknowledge support from SAO grant GO3-4016X and from the Swiss NSF (grant
20-66875.01). The MIT/HETGS group is also thanked 
for insightful information on off-axis spectroscopy with 
\chan.

\end{acknowledgements}

\end{document}